\documentclass[conference]{IEEEtran}

\usepackage{cite}
\usepackage{amsmath,amssymb,amsfonts}
\usepackage{algorithmic}
\usepackage{graphicx}
\usepackage{textcomp}
\usepackage{xcolor}
\def\BibTeX{{\rm B\kern-.05em{\sc i\kern-.025em b}\kern-.08em
    T\kern-.1667em\lower.7ex\hbox{E}\kern-.125emX}}

\usepackage{amsmath,graphicx,subcaption,caption,tabularx,bm}
\usepackage{comment,tikz,xcolor}
\usepackage{siunitx,amssymb}

\usepackage{balance}
\usepackage{flushend}

\usepackage{acronym}
\acrodef{IVAS}{Immersive Voice and Audio Services}
\acrodef{EVS}{Enhanced Voice Services}
\acrodef{3GPP}{Third Generation Partnership Project}
\acrodef{ISM}{Independent Streams with Metadata}
\acrodef{ParamISM}{Parametric ISM}
\acrodef{EFAP}{Edge Fading Amplitude Panning}
\acrodef{VBAP}{Vector-Base Amplitude Panning}
\acrodef{DTX}{Discontinuous Transmission}
\acrodef{MUSHRA}{Multiple Stimuli with Hidden Reference and Anchor}
\acrodef{WMOPS}{Weighted Million Operations Per Second}

\newsavebox{\mybox}

\begin{document}

\title{Parametric Object Coding in IVAS: Efficient Coding of Multiple Audio Objects at Low Bit Rates
}

\author{
\IEEEauthorblockN{Andrea Eichenseer\IEEEauthorrefmark{1},
Srikanth Korse\IEEEauthorrefmark{2},
Guillaume Fuchs\IEEEauthorrefmark{1}, and 
Markus Multrus\IEEEauthorrefmark{1}}
\IEEEauthorblockA{\IEEEauthorrefmark{1}Fraunhofer IIS, Erlangen, Germany, Email: {\{andrea.eichenseer, guillaume.fuchs, markus.multrus\}@iis.fraunhofer.de}}
\IEEEauthorblockA{\IEEEauthorrefmark{2}International Audio Laboratories Erlangen, Erlangen, Germany, Email: srikanth.korse@audiolabs-erlangen.de}
}

\maketitle

\begin{abstract}
The recently standardized 3GPP codec for Immersive Voice and Audio Services (IVAS) includes a parametric mode for efficiently coding multiple audio objects at low bit rates. In this mode, parametric side information is obtained from both the object metadata and the input audio objects. 
The side information comprises directional information, indices of two dominant objects, and the power ratio between these two dominant objects.
It is transmitted to the decoder along with a stereo downmix. 
In IVAS, parametric object coding allows for transmitting three or four arbitrarily placed objects at bit rates of 24.4 or 32 kbit/s and faithfully reconstructing the spatial image of the original audio scene. Subjective listening tests confirm that IVAS provides a comparable immersive experience at lower bit rate and complexity compared to coding the audio objects independently using Enhanced Voice Services (EVS).  
\end{abstract}


\acresetall
\section{Introduction}

The \ac{3GPP} recently standardized \ac{IVAS}~\cite{3GPP250,3GPP253,Multrus2024IVAS} with the aim of providing an immersive experience for several communication scenarios. It extends the \ac{3GPP} \ac{EVS} mono codec~\cite{3GPP445,Dietz2015} towards a multitude of audio formats such as stereo, multi-channel audio, scene-based audio (Ambisonics), and object-based audio.

One of the use cases of the IVAS object-based audio mode, referred to as \ac{ISM} mode, is multi-party conferencing~\cite{Fotopoulou2024IVAS} where multiple participants are combined into an audio scene. The participants might be located at the same or different locations, and the scene might be combined with corresponding video streams or avatars. In order to create an immersive experience, each audio object is associated with metadata corresponding to the real or virtual scene. This metadata consists of direction information, distance, gain, and, optionally, additional parameters.

In mobile communication, codecs have to operate at low bit rates and be efficient in terms of computational complexity.
Parametric coding approaches are especially suited for such bandwidth-constrained environments.
Several parametric methods have been proposed over the years to facilitate the transmission of spatial audio at low bit rates. Binaural Cue Coding~\cite{Faller2002_BCC,Faller2003_BCC_II,Baumgarte2003_BCC_I} was proposed to transmit multi-channel audio at lower bit rates by computing a single-channel downmix along with side information. While the side information mainly consists of interchannel cues, the mono downmix is computed by adding all the objects. Spatial Audio Object Coding (SAOC)~\cite{breebaart2008spatial, herre2012mpeg} enables the transmission of multiple objects at lower bit rates by transmitting either a mono or stereo downmix along with side information such as object level differences, inter-object cross coherence, downmix channel level differences, downmix gains, and object energies. At the decoder, the spatial audio scene is either rendered to stereo, binaural, or 5.1 loudspeaker format. 
Spatial Audio Object Coding 3D~\cite{murtaza2015iso_mpeg_h} extends SAOC by relaxing the constraint on the number of downmix channels. In addition, it also enables rendering to beyond the 5.1 loudspeaker layout.    
Joint Object Coding and Advanced Joint Object Coding~\cite{purnhagen2016immersive,Kjörling2014_AC4} transmit a downmix and metadata related to the downmix. In addition, another set of side information corresponding to the input audio objects and downmix is generated and transmitted. At the decoder, this side information is used to obtain the audio objects from the downmix signals. Finally, the decoded audio objects are rendered to a specific output layout using the metadata corresponding to the input objects.  

The IVAS codec is designed for mobile communication and thus follows strong requirements for delay, bit rate and complexity. For any immersive mode, a delay not exceeding $40$~\unit{ms} was mandated. Current mobile networks typically operate at bit rates not exceeding $24.4$~\unit{kbit/s} for voice communication. For efficiently coding multiple audio objects at such low total bit rates, strict requirements had to be set on the object side information: 
The side information bit rate of the previously mentioned methods is highly dependent on the number of objects and scales with an increase of the number of objects~\cite{breebaart2008spatial, herre2012mpeg, murtaza2015iso_mpeg_h, purnhagen2016immersive, Kjörling2014_AC4, Terentiev2009SAOC}. While this was acceptable for the broadcasting application they were designed for, using these methods for mobile communication at target bit rates of $24.4$ and $32$~\unit{kbit/s} would have been unfeasible.
In addition, the strong requirements on complexity not only include run-time complexity such as \ac{WMOPS}~\cite{ITU-T_G.191} and RAM demand, but also overall code size and ROM memory. To satisfy these requirements, an essential goal when designing the object-based audio mode within IVAS was the harmonization of components such as filter banks and rendering schemes across different coding modes.

In this paper, we present \ac{ParamISM}, the parametric object coding mode within \ac{IVAS}, which enables efficient coding of multiple audio objects at low bit rates. In particular, \ac{ParamISM} is used in \ac{IVAS} for efficiently coding three or four objects at $24.4$ or $32$~\unit{kbit/s}.
It integrates seamlessly into IVAS by sharing filter banks, core coding, and rendering schemes with other modes.
We show through subjective listening tests that, compared to independent coding of objects using \ac{EVS}, \ac{ParamISM} provides a better immersive experience at lower bit rates and complexity.

Sections~\ref{sec:enc} and \ref{sec:dec} give detailed descriptions of the encoder and decoder processing of ParamISM.
In Section~\ref{sec:evaluation}, a subjective and objective evaluation of \ac{ParamISM} against \ac{EVS} is provided and Section~\ref{sec:conclusion} concludes the paper.

\begin{figure*}[t]
    \centering
    
    \sbox{\mybox}{%
        \includegraphics[width=0.495\textwidth]{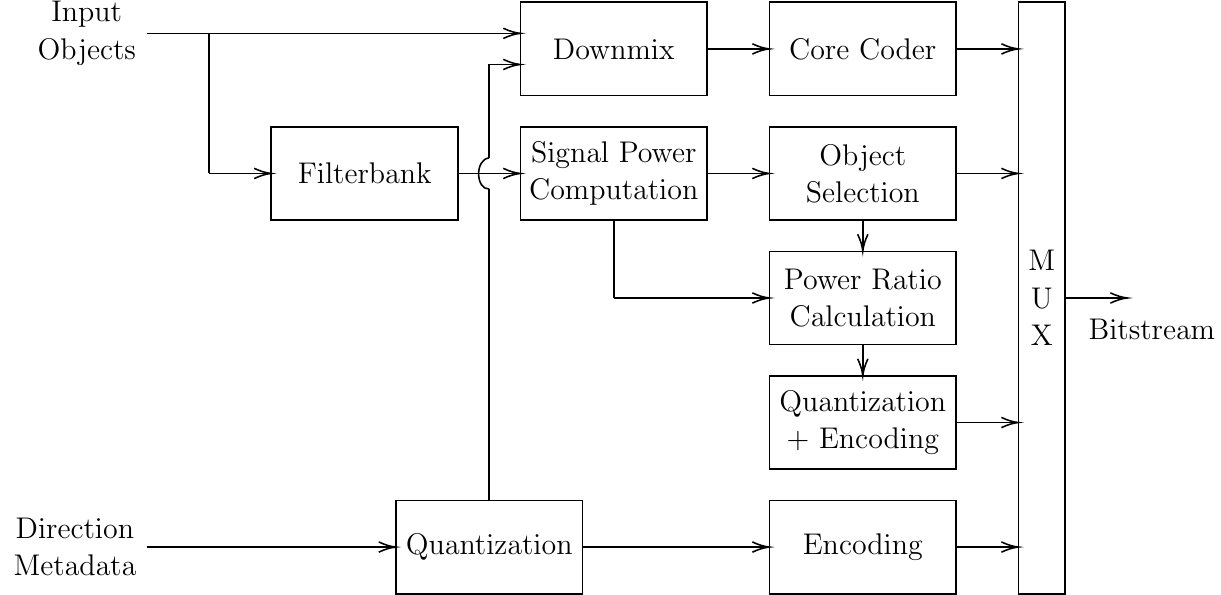}%
    }
    \begin{subfigure}[t]{0.495\textwidth}
        \centering
        \usebox{\mybox}
        \caption{ParamISM encoder.}
        \label{fig:paramism_enc}
    \end{subfigure}
    \hfill
    \begin{subfigure}[t]{0.495\textwidth}
        \centering
        \includegraphics[height=\ht\mybox]{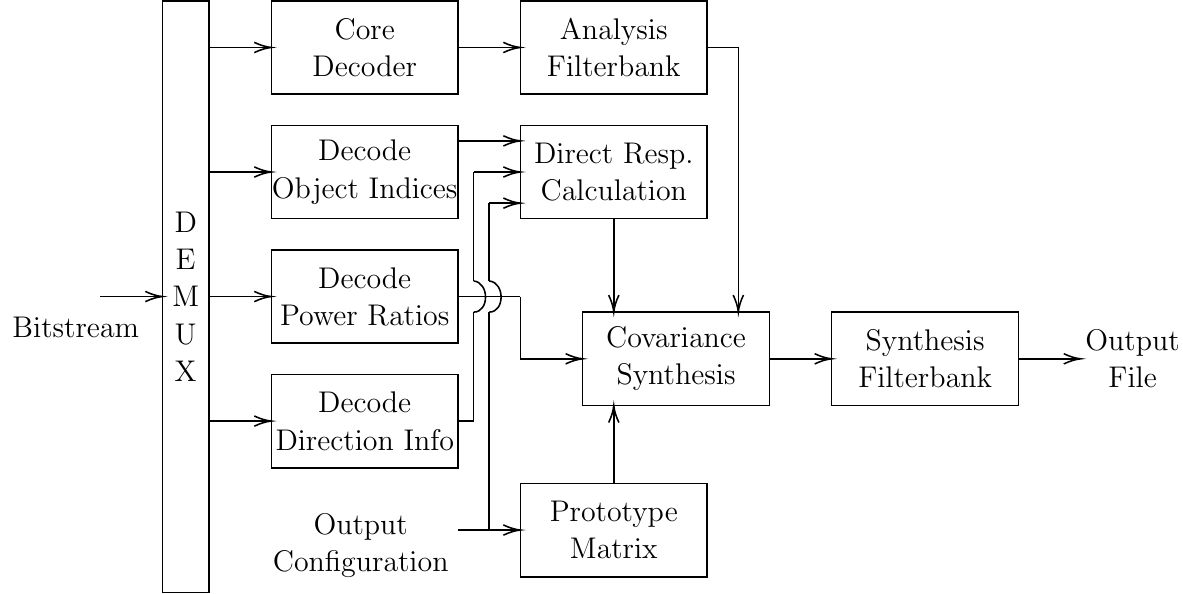}
        \caption{ParamISM decoder.}
        \label{fig:paramism_dec}
    \end{subfigure}
    
    \caption{Block diagrams of ParamISM encoder and decoder.}
\end{figure*}

\section{Parametric ISM Encoder}  
\label{sec:enc}

The presented parametric object coding scheme works in the time-frequency domain, where it is assumed that per each time-frequency tile, 
two most dominant objects can be identified based on which a power ratio between these two dominant objects is calculated and transmitted to the decoder along with the direction information of all objects.
This idea is derived from Directional Audio Coding (DirAC)~\cite{Pulkki2009DirAC}, where the side information is based on one directional component per time-frequency tile.
Our experiments showed, however, that for our purposes, two such directional components, or dominant objects, yield much better results.
Since the dominant objects typically vary across the time-frequency tiles due to different frequency content, enough spatial information is retained in the side information to ensure a faithful reconstruction of three or four objects.
The input audio objects themselves are transmitted by means of a two-channel time-domain downmix signal.
Fig.~\ref{fig:paramism_enc} shows the encoder block diagram with the main functional blocks.
In the following, we describe these functional blocks and give details on how the side information and downmix signal are obtained.

\subsection{Parametric Side Information}

The mono input objects $x_i$ are transformed into a time-frequency representation $X_i(k,n)$ by means of a filterbank, 
where $k$ and $n$ denote frequency bin and time slot indices, respectively. In IVAS, we use a Modified Discrete Fourier Transform analysis filterbank as defined in~\cite{3GPP253} 
which converts $20$~\unit{ms} frames into $4$ time slots and $240$ frequency bins at a sampling rate of $48$~\unit{kHz}.
For each time-frequency tile $(k,n)$, the signal power of the $i$-th object is then determined as $P_i(k,n) = |X_i(k,n)|^2$.
To reduce the data load, 
the $4$ time slots are grouped into one slot and the $240$ bins are grouped into $11$ psychoacoustically designed bands.
The parameter band borders are given by $B(l)$ and are specific to the input sampling rate.
The signal power values per parameter band $l$ are then summed according to:
\begin{equation}
    P_i(l) = \sum_n \sum_{k=B(l)}^{B(l+1)} P_i(k,n)
\end{equation}


For each parameter band, the two objects with the greatest power values are identified and a power ratio between them is determined:
\begin{equation}
    r_1(l) = P_1(l)/(P_1(l)+P_2(l))
\end{equation}
Note that $P_1(l)$ denotes the power of the most dominant object within the parameter band and $P_1(l) \geq P_2(l)$.
As both ratios sum to one, $r_2(l)$ is obtained by $r_2(l) = 1 - r_1(l)$ and does not need to be transmitted. $r_1(l)$ is quantized with three bits using:
\begin{equation}
r_{\text{idx}}(l) = \lfloor 2(r_1(l)-0.5)(2^3-1) \rfloor
\end{equation}

With $L=11$ parameter bands, the side information bit expenditure for each frame of $20$ \unit{ms} amounts to $4L$ bits for the dominant object identifications--each value is expressed with $2$ bits--and $3L$ bits for the power ratio indices. Once per frame, the quantized azimuth ($7$ bits) and elevation ($6$ bits) information of each object is also transmitted.

\subsection{Stereo Downmix}

The input audio objects $x_i$ are downmixed into two downmix channels using 
downmix gains that are derived from two cardioids and depend on the object positions--more specifically, the azimuth information $\theta_i$ of each object. The two cardioids are fixed and oriented towards 90° (left) and -90° (right), respectively. The downmix gains $w_{L,i}$ and $w_{R,i}$ for the left and right channel are given by: 
\begin{align}
    w_{L,i} &= 0.5 + 0.5 \cos\left(\theta_i - \pi/2\right)\\
    w_{R,i} &= 0.5 + 0.5 \cos\left(\theta_i + \pi/2\right) = 1 - w_{L,i}
\end{align}
The downmix channels are then obtained by:
\begin{equation}
    DMX_L = \sum_i x_i w_{L,i}\:,\:\:\: DMX_R = \sum_i x_i w_{R,i}    
\end{equation}
By taking into account the positions of the objects, all objects located in the left hemisphere are assigned a stronger weight for the left channel, while objects located in the right hemisphere are assigned a stronger weight for the right channel.
This facilitates the rendering process.
Our experiments further showed that one downmix channel is not enough to produce satisfactory results.

To prevent discontinuities, smoothing is applied between frames, and an energy compensation is conducted to account for differences between the power of the two dominant objects and the total power of all objects. 
The left and right downmix channels are 
encoded independently and transmitted to the decoder.
Details on the core coding aspects are outside the scope of this paper and can be found in~\cite{3GPP253}.

\begin{table*}[!t]
    \centering
    \caption{Details on the audio scenes included in the test set. Azimuth and elevation are given in degrees, where positive values denote the left/upper hemisphere and negative values denote the right/lower hemisphere.}
    \begin{tabularx}{\textwidth}{XXX@{\hspace{40pt}}X@{\hspace{20pt}}X}
        \hline
        Audio Scene & Type & Azimuth (deg) & Elevation (deg) & Motion \\
        \hline
        i1 & speech & 0, -180, 90, -90 & -5, -5, -5, -5 & static \\
        i2 & speech, female & 60, 30, -30, -60 & 0, 0, 0, 0 & static \\
        i3 & speech & 0, 60, -30 to -150, -45 & 0, 0, 15, 0 & object 3 \\
        i4 & speech, male & 75, 25, -25, -75 & 30, 30, 30, 30 & static \\
        i5 & speech & 45, -135, -45, 135 & 0, 0, 0, 0 & static \\
        i6 & music & 100, 10, -80, -170 & 50, 50, 50, 50 & static \\
        i7 & music & 90, 30, -30, -90 & 0, 0, 0, 0 & static \\
        i8 & music, vocals & 0, -70, 50, -20 & 20, 22, 18, 0 & static \\
        i9 & music & 20, 40, -30, -45 & -5, 20, 5, -5 & static \\
        i10 & mixed & 60 to -10, -60, 90, -90 & 19, 25, -15, 0 & object 1 \\
        i11 & vocals & 10, -10, 10, -10 & 15, 20, 30, 30 & static \\
        i12 & speech & 20, 20, -20, -20 & 0, 40, 0, 40 & static \\
        \hline
    \end{tabularx}
    \label{tab:testset}
\end{table*}

\section{Parametric ISM Decoder}
\label{sec:dec}

The \ac{IVAS} decoder receives a bitstream from which it extracts and decodes the transmitted downmix channels as well as the parametric side information.
This information is then used for rendering the downmix signals to the desired output format.
Fig.\ref{fig:paramism_dec} shows the decoder block diagram which we describe in the following.

\subsection{Parametric Side Information and Downmix}

The power ratios are obtained from the transmitted power ratio index:
\begin{equation}
    \hat r_1(l) = r_{\text{idx}}(l)/(2(2^3-1)) + 0.5\:,\:\:\: \hat r_2(l) = 1 - \hat r_1(l)
\end{equation}
Together with the decoded direction information and the object indices of the two most dominant objects per parameter band, the output signals can be created from the two decoded downmix channels.
For this, the downmix is first transformed into a time-frequency representation by means of a complex low-delay filterbank~\cite{Schnell2007CLDFB} which converts frames of $20$\,\unit{ms} into $16$ time slots and $60$ frequency bins. The analysis filterbank at the decoder differs from the analysis filterbank at the encoder to allow for a consistent rendering across different IVAS modes.

\subsection{Rendering}


\ac{IVAS} supports, among other output formats, the four loudspeaker layouts 5.1, 5.1.4, 7.1, and 7.1.4 according to the CICP setups as specified in~\cite{ISOIEC23091-3:2018}. 
To render the downmix to a specific loudspeaker layout, covariance synthesis~\cite{vilkamo2013optimized} is employed where a mixing matrix $\mathbf{M}$ is computed per time-frequency tile which transforms the transmitted downmix signals into the loudspeaker output signals
\begin{equation}
    \bm{y} = \mathbf{M}\bm{x}\:,
\end{equation}
where $\bm{x} = [X_1(k,n),X_2(k,n)]^T$ denotes the decoded downmix signals, $\bm{y} = [Y_1(k,n),Y_2(k,n), \ldots ,Y_S(k,n)]^T$ denotes the loudspeaker signals, and $S$ is the number of loudspeakers. 


In order to compute the mixing mixing matrix $\mathbf{M}$, the covariance synthesis requires an input covariance matrix $\mathbf{C}_x$, a target covariance matrix $\mathbf{C}_y$, and a prototype matrix $Q$. The prototype matrix $Q$ is dependent on the configured loudspeaker layout and is computed by mapping $DMX_L$ to all left-hemisphere loudspeakers and $DMX_R$ to the right-hemisphere loudspeakers. The center speaker is assigned both channels with a weight of $0.5$.

Since the input objects are uncorrelated, the input covariance matrix $\mathbf{C}_x$ is computed by considering only the main diagonal elements. It simplifies to:
\begin{equation}
    \mathbf{C}_x(k,n) =
    \begin{bmatrix}
    X_1(k,n)X_1(k,n)^* & 0 \\
    0 & X_2(k,n)X_2(k,n)^*
    \end{bmatrix}
\end{equation}
In order to compute the target covariance matrix $\mathbf{C}_y$, we need panning gains (also referred to as direct responses $dr$), a reference power $P_{DMX}$, and the decoded power ratios $\hat r_i$ which were calculated per parameter band $b$ and are now valid for all frequency bins contained in $b$. The target covariance matrix $\mathbf{C}_y$ is given by
\begin{equation}
    \mathbf{C}_y = \mathbf{R}\mathbf{E}\mathbf{R}^H\:,
\end{equation}
where $\mathbf{R}=[\bm{dr}_1, \bm{dr}_2]$ denotes the panning gains corresponding to the two dominant objects and $\mathbf{E}$ is a diagonal matrix with element $e_{i,i} = E_i = DP_i(k)$ containing the power corresponding to the $i$-th dominant object. It is given by
\begin{equation}
    DP_i(k) = \hat r_i\cdot P_{DMX}(k),\: i\in\{1,2\}\:,
\end{equation}
where $P_{DMX}(k) = \sum_{i=1}^2\sum_{n=0}^{15} |X_i(k,n)|^2$. 
The panning gains are computed for the two dominant objects using \ac{EFAP}~\cite{Borß2014EFAP}, an improved panning method compared to \ac{VBAP}~\cite{pulkki1997VBAP}, based on their corresponding direction information and the target loudspeaker layout.
A synthesis filterbank is applied on the output signals $\bm{y}$ to obtain the time-domain loudspeaker signals for playback.

\section{Evaluation}
\label{sec:evaluation}

To evaluate the presented parametric object coding scheme, we conducted subjective listening tests as well as objective complexity measurements.
For this, \ac{IVAS} is used in \ac{ParamISM} mode 
with three and four input objects plus associated metadata and compared against its predecessor \ac{EVS}.
As \ac{EVS} is a mono codec that is only able to process one monaural signal at a time, one instance of \ac{EVS} is executed per each input object.
We refer to this as multi-mono \ac{EVS}.

\subsection{Test Set}

A set of twelve audio scenes containing different kinds of audio objects is used for the subjective evaluation of the proposed method.
Since multi-party conferencing is probably the most relevant use case, most of the audio scenes feature speech uttered by different participants.
The amount of overlapped speech varies from scene to scene, but mostly, it is kept small to create realistic scenarios of polite discussions.
\ac{IVAS} also supports music and mixed audio, which is why some audio scenes contain general audio content in the form of music (different instruments and vocals) and ambient sound.

Table~\ref{tab:testset} summarizes the details of each audio scene of the test set.
Azimuth and elevation values are given for each of the $4$ objects of the scene; in case of a present moving object, the start and end position is given and the motion itself is uniform across all frames of the audio file.
The duration of the test files varies between $10$ and $21$ seconds at $48$~\unit{kHz}.

\begin{figure*}[t]
    \centering
    \begin{subfigure}[t]{0.495\textwidth}
        \centering
        \includegraphics[width=\linewidth]{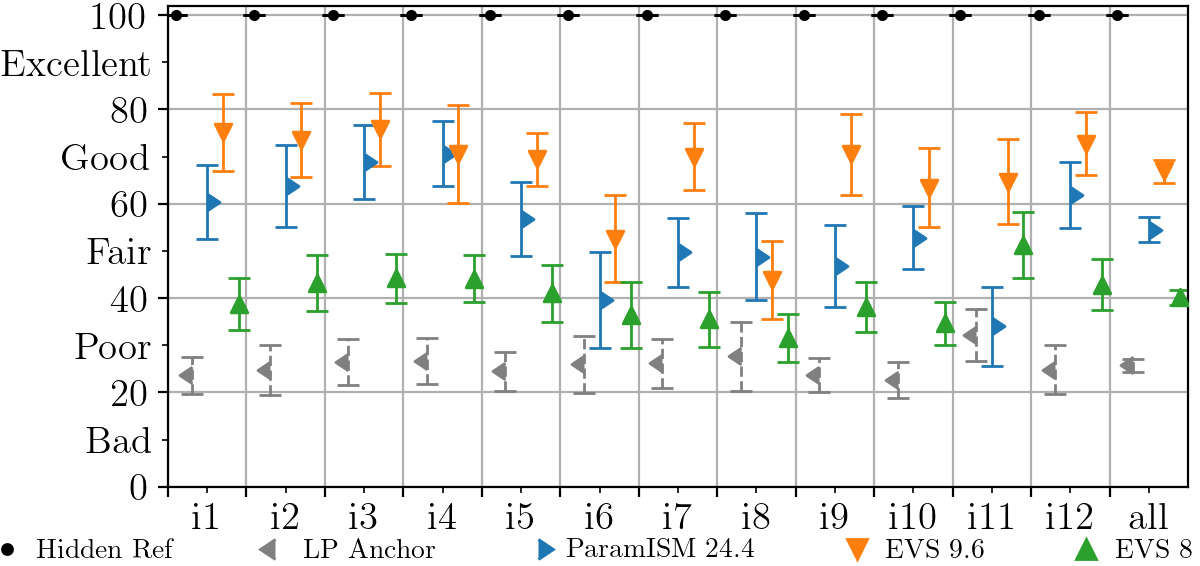}
        \caption{MUSHRA results for the 3-object listening test. IVAS at $24.4$~\unit{kbit/s} is compared against $3$x EVS at $8$~\unit{kbit/s} and $9.6$~\unit{kbit/s}.}
        \label{fig:3obj_MUSHRA}
    \end{subfigure}
    \hfill
    \begin{subfigure}[t]{0.495\textwidth}
        \centering
        \includegraphics[width=\linewidth]{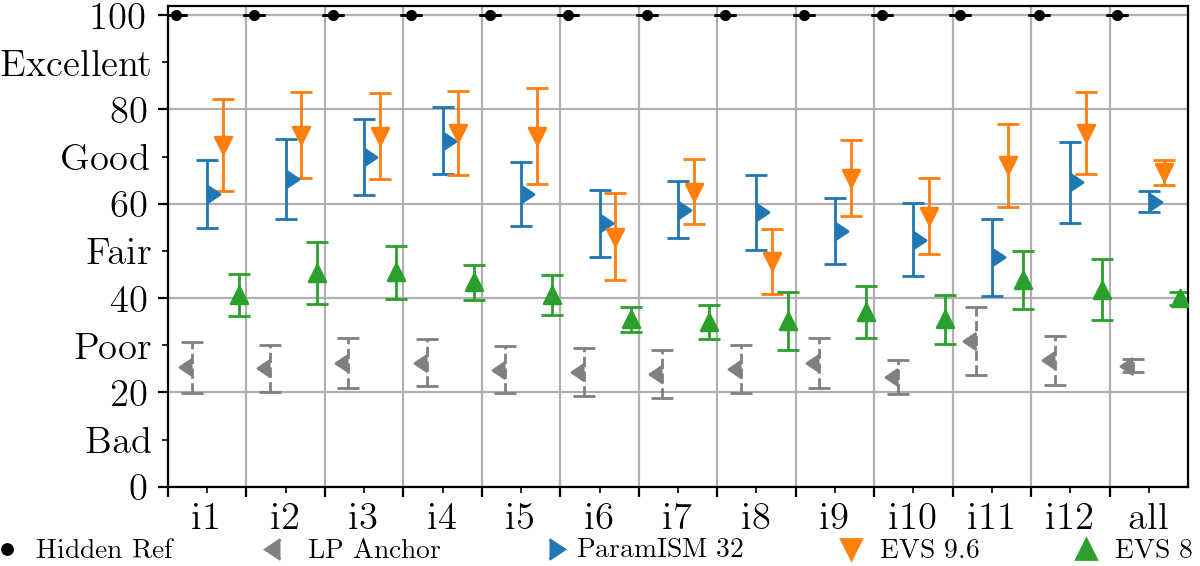}
        \caption{MUSHRA results for the 4-object listening test. IVAS at $32$~\unit{kbit/s} is compared against $4$x EVS at $8$~\unit{kbit/s} and $9.6$~\unit{kbit/s}.}
        \label{fig:4obj_MUSHRA}
    \end{subfigure}
    \caption{Listening test results. (a) Three input objects. (b) Four input objects.}
    \label{fig:MUSHRA}
\end{figure*}

\subsection{Subjective Test}

With the test set, two subjective listening tests following the \ac{MUSHRA} methodology~\cite{ITURBS1534} were conducted on a 7.1.4 loudspeaker layout.
The first test focuses on audio scenes containing three objects, while the second test focuses on scenes with four objects.
In the three-object test, the fourth objects of the scenes are not used.
The test comprises the following conditions: Hidden reference, 3.5~\unit{kHz} low-pass anchor, ParamISM at 24.4~\unit{kbit/s},  multi-mono EVS at 9.6~\unit{kbit/s} per object, and multi-mono EVS at 8~\unit{kbit/s} per object. For the four-object test, ParamISM at 32~\unit{kbit/s} is used instead of 24~\unit{kbit/s}. The rest of the conditions remain the same.

For multi-mono EVS, the coded objects were rendered to the loudspeaker layout using the unquantized object metadata in a post-processing step.
The gains for the rendering were computed using \ac{EFAP}~\cite{Borß2014EFAP}. 
In this context, it is important to note that the metadata bit rate is part of the total bit rate in \ac{IVAS}.
In contrast, \ac{EVS} uses the full bit rate for coding the objects without taking into account any metadata bit rate.
This puts multi-mono \ac{EVS} at an advantage over \ac{IVAS} when the exact same bit rates are compared.
On average, the metadata consumes $1.6$~\unit{kbit/s} per object in IVAS.

Figure~\ref{fig:MUSHRA} shows the results of the listening tests.
A total of $13$ expert listeners participated in both tests.
The results show that \ac{ParamISM} in \ac{IVAS} achieves an audio quality in the \textit{good} range for speech and in the \textit{fair} range for music and other audio content.
On average, it outperforms \ac{EVS} at the same bit rate by $15$ to $20$ MUSHRA points.
Notably, \ac{IVAS} \ac{ParamISM} supports super-wideband coding at the investigated bit rates, whereas multi-mono \ac{EVS} at $8$~\unit{kbit/s} per object only supports wideband coding.
At $9.6$~\unit{kbit/s} per object, EVS also supports super-wideband coding.
Especially in the four-object test, it is seen that ParamISM approaches the audio quality of EVS at the higher bit rate, while also including the coded metadata in the total rate.
Subtracting the metadata bit rate, ParamISM indeed uses only $61$\% of the rate of $4$x EVS at $9.6$~\unit{kbit/s} for coding the audio objects.
In the three-object case, ParamISM uses $63$\% of the rate of $3$x EVS at $9.6$~\unit{kbit/s}.

The quality of ParamISM drops somewhat in audio scenes that contain strong object overlaps, e.g., audio scenes i1 and i5 where there is a 50-100\% overlap of the talkers. One can argue, though, that such an overlap would not occur in a realistic conferencing or conversation scenario.
Scene i11 contains a recording of a choir, where, in addition to a 100\% temporal overlap, there is also the challenge of a large overlap in the frequency components.
Scene i6 proves somewhat challenging due to simultaneously occurring transients in all objects.
For pure music content, a higher bit rate is required to achieve \textit{good} to \textit{excellent} quality and for this, IVAS provides multiple other modes in the form of multi-channel and Ambisonics coding or non-parametric object coding at bit rates of up to $512$~\unit{kbit/s}.
However, ParamISM is still able to produce acceptable quality at low bit rates, retaining immersion for any spatial audio scene, while at the same time fulfilling the low-delay and complexity requirements of a communication codec.

\subsection{Complexity}

To also evaluate ParamISM in an objective manner, Table~\ref{tab:complexity} summarizes the complexity measured~\cite{ITU-T_G.191} in WMOPS for both \ac{IVAS} \ac{ParamISM} and multi-mono \ac{EVS}.
Measurements are conducted for ParamISM with three and four input objects.
Accordingly, three and four instances of EVS are run and measured for comparison.
The numbers show that the parametric object coding mode in IVAS generally requires fewer WMOPS to encode and decode audio scenes with three or four objects than multi-mono EVS.

\begin{table}[t]
    \centering    
    \caption{Complexity of IVAS ParamISM and multi-mono EVS for three and four objects measured in WMOPS. Numbers reported for EVS correspond to a summation over three or four instances.}
    \begin{tabular}{ lccc }
        \hline
         Condition, 3 objects & Encoder & Decoder & Total  \\
         \hline
         ParamISM at 24.4\,\unit{kbit/s} & 151.39 & 89.75 & 241.14 \\
         ParamISM at 32\,\unit{kbit/s} & \textbf{122.09} &	\textbf{75.53} &	\textbf{197.62} \\
         \text{3x EVS at 8\,\unit{kbit/s}} & 144.10 & 93.47 & 237.57 \\
         \text{3x EVS at 9.6\,\unit{kbit/s}} & 208.84 & 104.87 & 313.71 \\
         \hline
        Condition, 4 objects & Encoder & Decoder & Total  \\
         \hline
         ParamISM at 24.4\,\unit{kbit/s} & 153.99	& 94.18 & 248.18 \\
         ParamISM at 32\,\unit{kbit/s} & \textbf{124.95} &	\textbf{80.35} & \textbf{205.30} \\
         \text{4x EVS at 8\,\unit{kbit/s}} & 192.17 & 124.53 & 316.70 \\
         \text{4x EVS at 9.6\,\unit{kbit/s}} & 278.45 & 139.41 & 417.86 \\
         \hline
    \end{tabular}
    \label{tab:complexity}
\end{table}

\section{Conclusion}
\label{sec:conclusion}
In this paper, the parametric object coding mode ParamISM as implemented within \ac{IVAS} was presented. It was shown that employing a stereo downmix along with parametric side information based on two dominant objects allows for efficiently coding and transmitting multiple audio objects at low bit rates and rendering an immersive audio experience.
Compared against multi-mono \ac{EVS}, which requires external rendering of the objects, \ac{ParamISM} provides an all-in-one solution for coding and rendering and is even able to retain a higher audio bandwidth.
Within the \ac{IVAS} standard, \ac{ParamISM} is an efficient means for handling multi-party conferencing scenarios in low-delay and low-bit-rate environments.


\section*{Acknowledgment}
The authors would like to thank Stefan Bayer, Jan Frederik Kiene, Jürgen Herre, Archit Tamarapu, and Oliver Thiergart for their contributions to this work.

\flushend
\bibliographystyle{IEEEtran}
\bibliography{refs}

\end{document}